\documentclass[lettersize,journal]{IEEEtran}

\usepackage{cite}
\usepackage{amsmath,amssymb,amsfonts}
\usepackage{algorithm,algpseudocode}
\usepackage{graphicx}
\usepackage{textcomp}
\usepackage[svgnames]{xcolor}
\usepackage{tabularx,subcaption}
\usepackage{booktabs, titlesec}
\begin{document}

\title{DRL-Based AoI Minimization for RSMA in Finite-Blocklength MU-MISO}
\author{Enes Kaya, Mehmet Alp Demircioğlu, Elif Tugce Ceran,
Melda~Yuksel,~\IEEEmembership{Senior Member,~IEEE}
\thanks{Authors are with the Department of Electrical and Electronics Engineering, Middle East Technical University, Ankara, 06800, Turkey, e-mail: \{enes.kaya\_01, alp.demircioglu, elifce, ymelda\}@metu.edu.tr. 

E. Kaya is also with Aselsan Inc., Ankara, 06200, Turkey, e-mail: eneskaya@aselsan.com.}}

\maketitle

\begin{abstract}

This letter investigates age-of-information (AoI) minimization in multi-user wireless networks operating in the finite-blocklength (FBL) regime, which is critical for low-latency transmission of short state-update packets.  While rate-splitting multiple access (RSMA) provides a powerful and flexible framework for interference management in multi-user FBL systems, the joint optimization of its parameters, such as precoding vectors, power allocation, and rate-splitting ratios, to guarantee information freshness results in analytically intractable complexity.
To address this challenge, we propose an actor--critic deep reinforcement learning (DRL) framework to learn dynamic resource-allocation policies in multi-user multiple-input single-output (MU-MISO) broadcast channels. Simulation results show that the proposed RSMA-RL framework achieves consistently lower AoI than the state-of-the-art benchmarks, with substantial gains observed at low signal-to-noise ratio (SNR) and short blocklengths, while matching benchmark performance at high SNR with significantly lower online complexity via a single neural-network forward pass at execution.

\end{abstract}


\section{Introduction}

Managing interference is a key challenge in 6G systems supporting massive connectivity under stringent reliability and latency requirements. Multi-user multiple-input multiple-output (MU-MIMO) improves spectral efficiency through spatial multiplexing, but conventional schemes struggle in dense and heterogeneous networks. Emerging holographic MIMO architectures provide additional spatial degrees of freedom for interference management and transmission optimization, further motivating flexible transmission strategies \cite{ICSaT_24_2022_Yijie,liu2017nonorthogonal,spherical2026}. Among these strategies, rate-splitting multiple access (RSMA) has emerged as a powerful method for interference management. In RSMA, each user's message is split into common and private parts, enabling partial interference decoding while treating the remainder as noise. This flexible approach bridges full interference decoding and treating interference entirely as noise, improving robustness and spectral efficiency \cite{ICSaT_24_2022_Yijie}. These advantages motivate our focus on this RSMA framework.

In addition to these advancements, the growing demand for 6G hyper-reliable and low-latency communications (HRLLC) has pushed finite-blocklength (FBL) analysis to the forefront. In this regime, short packets lead to a non-negligible decoding error probability, and classical Shannon capacity no longer fully captures performance. FBL models are therefore essential for understanding the reliability--latency tradeoffs that arise in mission-critical applications \cite{ITIT_56_2010_Yury,bennis2018ultrareliable}.

Latency alone does not fully capture information freshness. The age of information (AoI) metric \cite{kaul2012real} quantifies the freshness of status updates in applications such as IoT sensing, industrial automation, and real-time monitoring. Under FBL constraints, AoI captures a key tradeoff: shorter blocklengths reduce transmission time but increase the block error rate (BLER). Recent studies have applied RSMA to AoI optimization, demonstrating improved information freshness through better interference management in asymptotic regimes \cite{IWCL_13_2024_Onur,akyurek2023age}. However, the joint optimization of RSMA parameters for long-term AoI minimization remains analytically intractable in dynamic environments, while conventional methods require repeated optimization for each channel realization. This motivates the use of deep reinforcement learning (DRL), which can learn near-optimal resource-allocation policies from the current channel and AoI states without requiring an explicit channel-distribution model and enables low-complexity online inference \cite{IA_11_2023_Aysenur,Liu_2022,liu2023deep}. In particular, actor–critic methods, such as deep deterministic policy gradient (DDPG) \cite{lillicrap2019ddpg} and twin-delayed deep deterministic policy gradient (TD3) \cite{fujimoto2018td3}, are well suited due to their ability to efficiently handle the continuous action space of joint precoding and power allocation \cite{S_23_2023_Qiong,IWCL_11_2022_Xue,ITMC_23_2024_Xingqiu}.



Learning-based RSMA designs have been explored for maximization of the sum-rate in interference channels without considerations of AoI \cite{irkicatal2024deep}. Existing works on AoI optimization consider DRL-based approaches in single-antenna settings \cite{ITMC_24_2025_Congwei} or successive convex approximation (SCA)-based MU-MISO precoding \cite{ITWC_23_2024_Minsu}; however, they do not account for FBL communications, or do not incorporate rate-splitting.

AoI minimization in MU-MISO systems under FBL constraints with RSMA remains fundamentally unresolved. Existing work does not integrate precoding, power allocation, and multiple access design into a unified learning-based framework. Motivated by these gaps, this letter develops a learning-driven approach for AoI minimization in MU-MISO downlink under the FBL regime using RSMA. Our contributions are: (i) a novel unified framework that captures the key latency--reliability--freshness tradeoff achievable with RSMA; (ii) an actor-critic DRL architecture, implemented using DDPG and TD3, that jointly controls the common/private stream powers and precoders to minimize AoI; and (iii) significant gains over state-of-the-art RSMA benchmarks, particularly in challenging low signal-to-noise ratio (SNR) and short-blocklength regimes, achieved with only a single forward pass at execution.

To the best of our knowledge, this joint optimization problem is posed for the first time in the literature, and this letter presents the first systematic DRL framework for AoI minimization in multi-user finite-blocklength RSMA systems.

\section{System Model and Problem Formulation}\label{sec:sysmodel}

This work examines a downlink MU-MISO system, where the base station (BS) uses $M$ antennas to provide service to $K$ single-antenna users, represented by the index set $\mathcal{K} = \{1, 2, \dots, K\}$. We employ a 1-layer RSMA scheme, characterized by one common stream intended for all users and $K$ private streams, each exclusively dedicated to a specific user. For each user $k \in \mathcal{K}$, the BS holds a distinct message, $B_k$. Let $I_k = l_k/n$ be the total information rate required for transmitting message $B_k$ of length $l_k$ bits to user $k$. The number of bits $l_k$, and the packet size $n$ are assumed to be fixed for a given system. In the simulation results, we will study how AoI changes with variable $n$ for fixed $l_k$.

In accordance with the RSMA protocol, every message $B_k$ is partitioned into two components: a \textit{common part}, $B_{c,k}$, and a \textit{private part}, $B_{p,k}$. The collective set of common parts, $\{B_{c,1}, \dots, B_{c,K}\}$, is jointly encoded into a single common data stream, denoted as $b_c$. Each private part $B_{p,k}$ is independently encoded into its own private data stream, $b_k$. 


The resulting streams are aggregated into a single symbol vector, $\mathbf{b} = [b_c, b_1, \dots, b_K]^T$. This vector undergoes linear precoding by a matrix $\mathbf{P} = [\mathbf{p}_c, \mathbf{p}_1, \dots, \mathbf{p}_K]$, where $\mathbf{p}_j \in \mathbb{C}^{M \times 1}$ is the designated precoding vector for stream $b_j$. The final signal transmitted from the BS is thus $\mathbf{x} = \mathbf{P} \mathbf{b} = \mathbf{p}_c b_c + \sum_{k \in \mathcal{K}} \mathbf{p}_k b_k.$
Assuming the transmitted symbols possess unit power, i.e., $\mathbb{E}[\mathbf{b}\mathbf{b}^H] = \mathbf{I}$, the total transmit power is constrained by a maximum value $P_{\text{max}}$
\begin{equation}
    \mathbb{E}[\|\mathbf{x}\|^2] = \mathrm{tr}(\mathbf{P} \mathbf{P}^H) \leq P_{\text{max}}. \label{eqn:powerconst}
\end{equation}
The received signal at user $k$ is modeled as $
    y_k = \mathbf{h}_k^H \mathbf{x} + z_k$, $\forall k \in \mathcal{K}$,
with $\mathbf{h}_k \in \mathbb{C}^{M \times 1}$ being the channel vector from the BS to user $k$, and $z_k \sim \mathcal{CN}(0,\sigma_k^2)$ denoting the receiver's circularly symmetric complex Gaussian noise. The quasi-static channel vectors remain constant for the duration of a single transmission block but vary independently from one block to the next \cite{yang2014quasi}. To decode their messages, each user implements  successive interference cancellation (SIC).
    
    \emph{Common Stream Decoding:} Initially, user $k$ decodes the common stream $b_c$, treating all private streams as interference. The signal-to-interference-plus-noise ratio (SINR) for this stage is $
        \gamma_{c,k} = | \mathbf{h}_k^H \mathbf{p}_c |^2 / \left(\sum_{j \in \mathcal{K}} | \mathbf{h}_k^H \mathbf{p}_j |^2 + \sigma_k^2\right).$
    

    \emph{Private Stream Decoding:} Upon successful decoding of $b_c$, its contribution is perfectly removed from the received signal $y_k$. User $k$ then proceeds to decode its own private stream $b_k$, treating the private streams of all other users ($j \neq k$) as residual interference. The SINR for this second stage is $
        \gamma_{p,k} = | \mathbf{h}_k^H \mathbf{p}_k |^2/ \left(\sum_{j \in \mathcal{K}, j \neq k} | \mathbf{h}_k^H \mathbf{p}_j |^2 + \sigma_k^2\right).$
Finally, user $k$ reconstructs $B_k$ by combining the decoded common part $B_{c,k}$ (extracted from $b_c$) with the decoded private part $B_{p,k}$.

The achievable rates are approximated using FBL theory. For a transmission blocklength of $n$  and target BLERs of $\epsilon_{c,k}$ and $\epsilon_{p,k}$, the rates for the common message and the private messages is given by $
    R_{i,k} \approx \log_2(1 + \gamma_{i,k}) - \sqrt{\frac{V_{i,k}}{n}} Q^{-1}(\epsilon_{i,k}), $
where $i \in \{c, p\}$ and $V_{i,k}$ denotes the channel dispersion for stream $i$ at user $k$, $V_{i,k} = \frac{\gamma_{i,k}(\gamma_{i,k} + 2)}{2(\gamma_{i,k} + 1)^2} \log_2^2(e)$.
The maximum decodable rate of the common stream, $R_c$, is limited by the weakest user and is given by $R_c = \min_{k \in \mathcal{K}} \{ R_{c,k} \}$.
Similarly, the private stream rate for user $k$ is expressed as $
R_{p,k} = r_k I_k,$
where $r_k \in [0,1]$ is the rate-splitting ratio, representing the fraction of the total information rate allocated to the private part. Consequently, the common portion requires a rate of $(1 - r_k) I_k$ for each user $k$ and the total common rate denoted by $R_c$ is given as $R_c = \sum_{k \in \mathcal{K}} (1 - r_k) I_k.$ Note that, the set of all rate-splitting ratios forms the rate-splitting vector, $\mathbf{r} \triangleq [r_1, r_2, \dots, r_K]^T$. 

By rearranging the FBL rate expressions, we can obtain the BLERs $\epsilon_{c,k}$ and $\epsilon_{p,k}$ as
\begin{align}
\epsilon_{c,k} &\approx Q\left( \frac{\log_2(1 + \gamma_{c,k}) - R_c}{ \frac{\log_2(e)}{\gamma_{c,k} + 1} \sqrt{ \frac{\gamma_{c,k}(\gamma_{c,k} + 2)}{2n} } } \right), \label{eqn:3}\\
\epsilon_{p,k} &\approx Q\left( \frac{\log_2(1 + \gamma_{p,k}) - R_{p,k}}{ \frac{\log_2(e)}{\gamma_{p,k} + 1} \sqrt{ \frac{\gamma_{p,k}(\gamma_{p,k} + 2)}{2n} } } \right).
\end{align}


\begin{figure*}[tb]
\centering
\includegraphics[width=0.65\textwidth]{"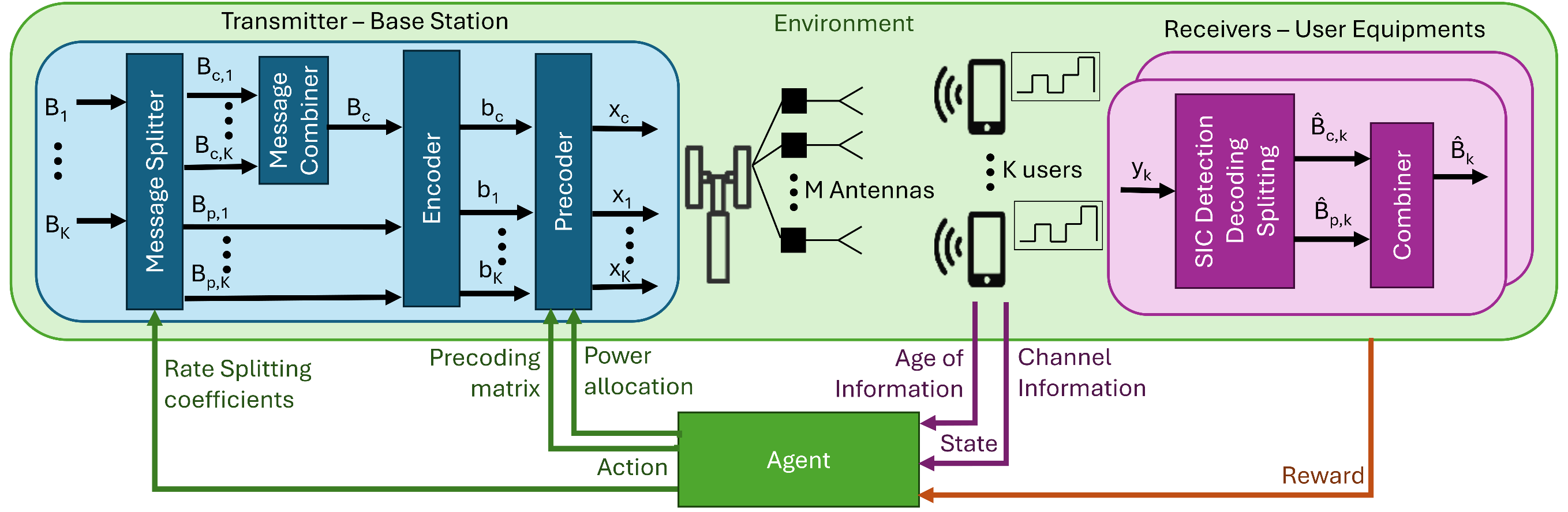"}
\caption{RSMA-based MU-MISO system with AoI-aware DRL resource allocation.}
\label{fig:agent}
\end{figure*}For user $k$, the overall BLER $\epsilon_k$ is the probability that either the common or the private stream is decoded in error, where 
$ \epsilon_k = 1 - (1 - \epsilon_{c,k})(1 - \epsilon_{p,k}) = \epsilon_{c,k} + \epsilon_{p,k} - \epsilon_{c,k} \epsilon_{p,k}.$ Note that, the selected common rate $R_c$ is reflected in the common-message BLER in \eqref{eqn:3}, with rates exceeding this limit resulting in increased decoding errors.

\subsection{AoI Model}

The AoI for any user $k$ at time $t$, denoted by $\Delta_k(t)$, represents the freshness of the user's information and is calculated as the elapsed time since the most recent update, which is denoted as $u_k(t)$, was generated; i.e., $\Delta_k(t) = t - u_k(t).$ Our analysis is based on a \textit{generate-at-will} model in slotted time, where each slot corresponds to one transmission block of length $n$. The AoI evolution for user $k$ depends on the decoding outcome. With probability $1-\epsilon_k$ the AoI is reset to $n$, while with probability $\epsilon_k$ it increases by $n$, yielding
\begin{equation}
    \Delta_{k}(t+1) =
    \begin{cases}
        n & \text{with probability } 1 - \epsilon_k \\
        \Delta_{k}(t) + n & \text{with probability } \epsilon_k
    \end{cases}.
\end{equation}

The objective is to minimize the long-term average AoI across all users by designing an optimal transmission policy that governs resource allocation; i.e., the rate splitting ratio vector $\mathbf{r}$, and the precoding matrix $\mathbf{P}$. The resulting optimization problem is defined as
\begin{equation}
    \min_{\mathbf{r}, \mathbf{P}} \lim_{T \to \infty} \frac{1}{T} \mathbb{E}\!\left[ \sum_{t=1}^{T} \sum_{k=1}^{K} \Delta_k(t) \right] \mbox{~subject~to \eqref{eqn:powerconst}}.
    \label{eq:AoI_obj}
\end{equation}



\section{DRL-Based AoI Minimization for FBL-RSMA}

In this section, we present a DRL framework for AoI minimization in FBL systems, formulated as a Markov decision process (MDP) and solved using two deterministic actor-critic algorithms, DDPG and TD3.

The problem is formulated as a discounted MDP, specified by the tuple $(\mathcal{S}, \mathcal{A}, \mathcal{P}, R, \gamma)$.  The state space $\mathcal{S}= (\mathbb{Z}^+)^K \times \mathcal{H}$ is continuous and high-dimensional, which comprises the current AoI of all users and the channel state information, $\mathbf{H}_t$. Here $\mathcal{H}$ is the set of all channel matrices $\mathbf{H}=[\mathbf{h}_1,...,\mathbf{h}_K]\in\mathbb{C}^{M\times K}$. At time $t$, the state is $s_t\triangleq(\Delta_1(t),\ldots,\Delta_K(t),\mathbf{H}_t)$. Action $a_t$ at time $t$ is defined as $a_t \triangleq (\mathbf{W}_t, \boldsymbol{\alpha}_t, \mathbf{r}_t)\in \mathcal{A}=\mathcal{W}\times\mathcal{P}\times\mathcal{R}$, where $\mathcal{W}$ denotes the set of normalized precoding matrices $\mathbf{W}=[\mathbf{w}_c,\mathbf{w}_1,\ldots,\mathbf{w}_K]$. To reduce the dimensionality of the action space, the normalized precoding vectors are parameterized in polar coordinates, and the corresponding angular parameters are learned by the RL agent. Thus, the unit-norm constraint $\|\mathbf{w}_j\|_2=1$ is satisfied by construction.
Similarly, the set $\mathcal{P}$ is the set of all power allocation coefficient vectors $\boldsymbol{\alpha} = [\alpha_c, \alpha_1, \dots, \alpha_K]^T$ that satisfy the total power constraint
 $\mathcal{P} = \left\{ \boldsymbol{\alpha} \in \mathbb{R}^{(K+1)\times1} \mid \|\boldsymbol{\alpha}\|_1 \leq P_{\text{max}} \right\}.$
    Note that the final precoding vectors are constructed as $\mathbf{p}_j = \sqrt{\alpha_j}\mathbf{w}_j$. Finally,  $\mathcal{R}$ is the set of all rate-splitting ratio vectors $\mathbf{r}$ such that
    $\mathcal{R} = \left\{ \mathbf{r} \in \mathbb{R}^{K\times1} \mid r_j \in [0,1] \text{ for } j \in \{1, \dots, K\} \right\}.$

State transitions follow directly from the AoI recursion, where decoding success or failure for each user is governed by the corresponding BLER $\epsilon_k(\mathbf{H}_t,a_t)$. The reward function $R: \mathcal{S} \times \mathcal{A} \rightarrow \mathbb{R}$ is defined as the negative sum of the next-state AoIs $R(s_t, a_t) = -\sum_{k=1}^{K}  \Delta_{k}(t+1).$

We employ DDPG and TD3 for AoI minimization in the high-dimensional continuous action space, using identical state-action representations, rewards, actor architectures, and constraint handling. The system model is depicted in Fig.~\ref{fig:agent}.






\noindent\textbf{State Representation:} The input to our networks is the state $s_t = (\boldsymbol{\Delta}_t, \mathbf{H}_t)$. To be processed by a neural network, this composite state is flattened into a single vector. The AoI vector $\boldsymbol{\Delta}_t \in (\mathbb{Z}^+)^K$ is concatenated with the real and imaginary parts of the flattened channel matrix $\mathbf{H}_t \in \mathbb{C}^{M \times K}$.

\noindent\textbf{Actor Network:} The actor network, $\mu(s | \theta^\mu)$, maps the input state vector to a deterministic action. Its output layer produces a raw action vector that is subsequently processed to conform to the constraints of the action space $\mathcal{A}$ as follows:

\noindent \textbf{Normalized Precoding Matrix $\mathbf{W}_t$:} A segment of the actor's raw output is reshaped into a complex-valued matrix of size $M \times (K+1)$. Each column $\mathbf{w}_j$ of this matrix is then normalized to ensure unit norm, i.e., $\mathbf{w}_j \leftarrow \mathbf{w}_j / \|\mathbf{w}_j\|_2$, thus forming the normalized precoding matrix $\mathbf{W}_t$.

\noindent \textbf{Power Allocation $\boldsymbol{\alpha}_t$:} To always use maximum power and have $\sum_{j \in \{c, 1, \dots, K\}} \alpha_j = P_{\text{max}}$. Thus, only $K$ variables are independent. The actor outputs a $K$-dimensional vector to determine the power distribution coefficients $\beta_j$. The final power allocation for each stream is then $\alpha_j = P_{\max} \cdot \beta_j$, $j=1,...,K$ and $\alpha_c$ is computed accordingly.

\noindent \textbf{Rate-Splitting Ratios $\mathbf{r}_t$:} The final segment of the actor's output is a $K$ dimensional vector, which naturally constrains each element $r_k$ to the interval $[0, 1]$.

\noindent\textbf{Critic Networks:} DDPG employs a single critic network
\(Q(s,a\mid\theta^Q)\), whereas TD3 employs twin critic networks,
\(Q_1(s,a\mid\theta^{Q_1})\) and
\(Q_2(s,a\mid\theta^{Q_2})\), to estimate the action-value function.  Each critic takes the flattened state and complete action vector $a_t=(\mathbf{W}_t,\boldsymbol{\alpha}_t,\mathbf{r}_t)$ as input and outputs a scalar Q-value. TD3 mitigates overestimation by using the minimum of the two critic estimates during updates.




At each time step, the actor selects an action with Gaussian exploration noise, executes it, and stores $(s_t,a_t,R_t,s_{t+1})$ in the replay buffer $\mathcal{B}$. The actor and critics are then updated using sampled transitions, with the critics minimizing the mean squared Bellman error via delayed target networks. TD3 follows \cite{fujimoto2018td3}, while DDPG uses a single critic without target-policy smoothing or delayed actor updates \cite{lillicrap2019ddpg}.

\noindent{\textbf{Complexity Analysis:}}
Both \textit{RSMA-LowLatency} \cite{ITVT_71_2022_Yunnuo} and \textit{RSMA-MaxMinFair} \cite{ITVT_72_2023_Yunnuo} rely on SCA-based algorithms with high computational complexity, given by $\mathcal{O}(\delta^{-1}\log(\tau^{-1})[KN_t]^{3.5})$, where $\delta$ denotes the step size.
For the proposed DRL-based algorithm, an actor network with $L^a$ layers and a critic network with $L^c$  layers incur computational costs given by $C_{\text{actor}} = \sum_{i=1}^{L^a} \mathcal{O}(n_{i-1}^a n_i^a)$ and $C_{\text{critic}} = \sum_{i=1}^{L^c} \mathcal{O}(n_{i-1}^c n_i^c)$. The overall training complexity is $\mathcal{O}(H U B (C_{\text{actor}} + C_{\text{critic}}))$, while inference requires only a single actor forward pass, $\mathcal{O}(C_{\text{actor}})$. In contrast to SCA-based methods that incur high complexity at each channel realization, the DRL-based approach shifts this burden to offline training, enabling significantly lower online computational complexity.

\section{Numerical Results and Discussion}



To optimize the performance of the DRL agents for the MU-MISO  RSMA setting, an exhaustive search over the hyperparameter space was conducted, the final hyperparameter set is detailed in Table \ref{table:hyper_rsma_mimo}. For the simulations, we consider the case where $M=2$ and $K=2$, providing a proof-of-concept evaluation. In the perfect CSIT case, the channels follow a memoryless quasi-static block-fading model with the entries in $\mathbf{H}_t$ independent circularly symmetric Gaussian, with zero mean and unit variance. There is additive white Gaussian noise, \(z_k \sim \mathcal{CN}(0,\sigma_k^2)\), with \(\sigma_k^2=1\) for all users. In the imperfect CSIT case, the channel follows a Gauss-Markov model: $\mathbf{H}_t=\xi \mathbf{H}_{t-1}+\mathbf{U}_t$, where $\xi\in [0,1]$ is the correlation coefficient between consecutive channel gains. The entries in $\mathbf{U}_t$s are independent circularly symmetric complex Gaussian with zero mean and variance 
$(1-\xi^2)$. Moreover, the base station only has access to previous channel realizations, and this knowledge
is also imprecise. Specifically, the base station is only aware of the first digit after the decimal point, lacking information on the subsequent fractional digits. In our RL methods, the agents are trained in perfect CSIT environment, then tested in imperfect CSIT. In the simulations $\xi=0.98$. In both cases SNR is also equal to \(P_{\max}\). Each status-update packet contains \(l_k=256\) information bits, and the AoI is initialized as \(\Delta_k(0)=1\), $\forall k$.

\begin{figure}[tb]
   \centerline{
   \includegraphics[width=0.45\textwidth]{"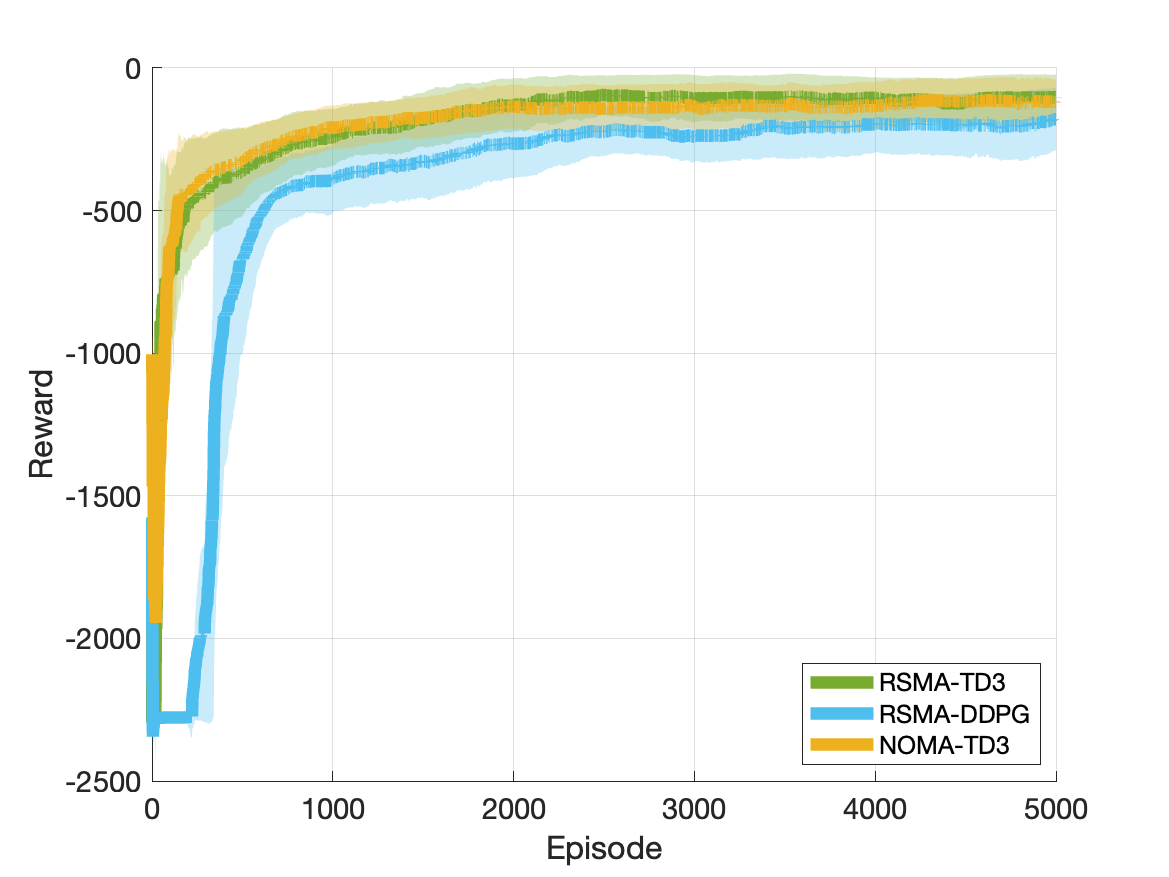"}}
   \caption{Training convergence of RSMA-TD3, RSMA-DDPG, and NOMA-TD3. Lines and shaded areas show the median and median absolute deviation over 500 episodes of 1000 steps.} 
   \label{fig:rsmatrain}
\end{figure}

Two optimization-based RSMA schemes are used as benchmarks. \textit{RSMA-LowLatency} \cite{ITVT_71_2022_Yunnuo}, employs a successive convex approximation (SCA)-based algorithm to minimize transmission delay under FBL constraints, while \textit{RSMA-MaxMinFair} \cite{ITVT_72_2023_Yunnuo} uses one-dimensional SCA to maximize max-min fairness, which is relevant to AoI minimization by preventing any user's AoI from dominating. Since no existing work addresses AoI minimization in this setting, both schemes are evaluated using the same FBL BLER and AoI model as defined in Section~\ref{sec:sysmodel}. We also consider a greedy AoI-based maximum-ratio transmission (MRT) without RSMA and an RL-based non-orthogonal multiple access (NOMA) scheme, which performs learning-based resource allocation under the same AoI objective. Fig.~\ref{fig:rsmatrain} presents the training reward curves of the proposed RSMA-RL agents, as well as for the NOMA-RL illustrating convergence and stability.

\begin{table}[t]
\centering
\caption{Hyperparameters of  DDPG- and TD3-Based RSMA}\vspace{-0.05in}
\label{table:hyper_rsma_mimo}
\setlength{\tabcolsep}{2.2pt} 
\begin{tabular}{l r l r}
\toprule
\textbf{Parameter} & \textbf{Value} & \textbf{Parameter} & \textbf{Value} \\
\midrule
Mini-Batch Size & 512 & 
Exp. Buffer Length & $1\times10^{6}$ \\
Target Smooth Fac. & $5\times10^{-3}$ & 
Target Update Freq. $d$ & 2 \\
Exploration Std Dev. & 0.3 & 
Exploration Std Decay & $1\times10^{-6}$ \\
Min Exploration Std. & 0.05 & 
Learning Rate & $3\times10^{-5}$ \\
Steps per Episode & 1000 & 
No of Fully Connected Layers & 3 \\
Optimizer & Adam & 
 Fully Connected Layer Size & 512\\
Gradient Threshold & 1 & 
L2 Regularization Fac. & $3\times10^{-6}$ \\
\bottomrule
\end{tabular}
\end{table}

\begin{figure*}
    \centering
    \begin{subfigure}{0.40\linewidth}
        \centering
        \includegraphics[width=\linewidth]{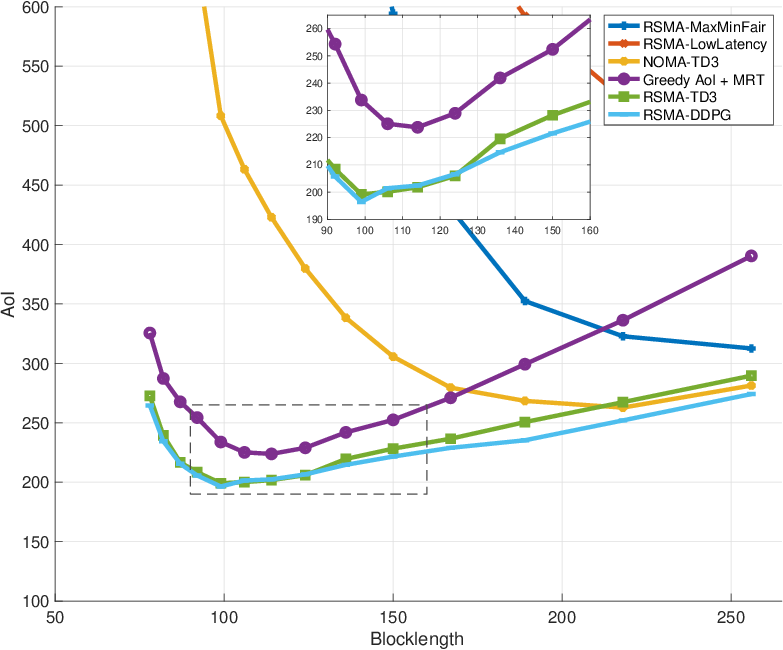}
        \caption{}
        \label{fig:a}
    \end{subfigure}
    \begin{subfigure}{0.40\linewidth}
        \centering
        \includegraphics[width=\linewidth]{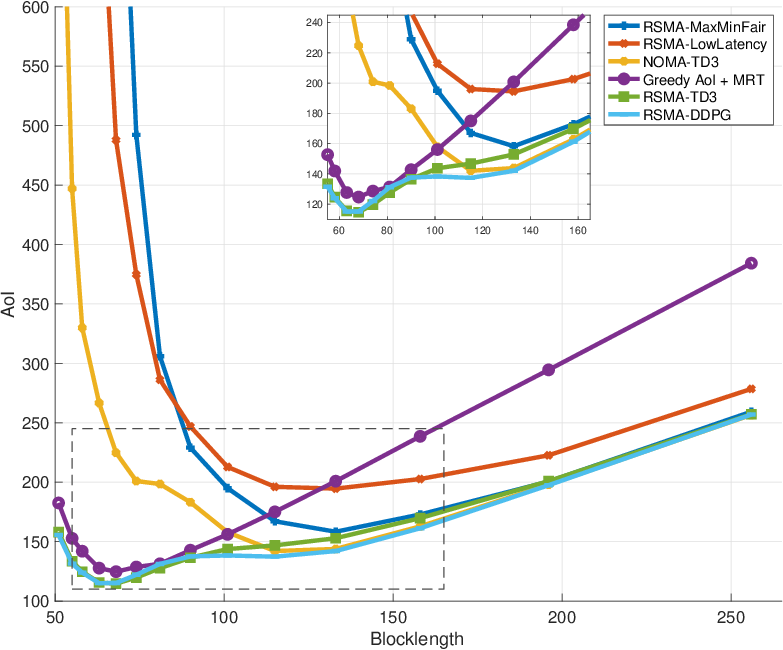}
        \caption{}
        \label{fig:b}
    \end{subfigure}
    \begin{subfigure}{0.40\linewidth}
        \centering
        \includegraphics[width=\linewidth]{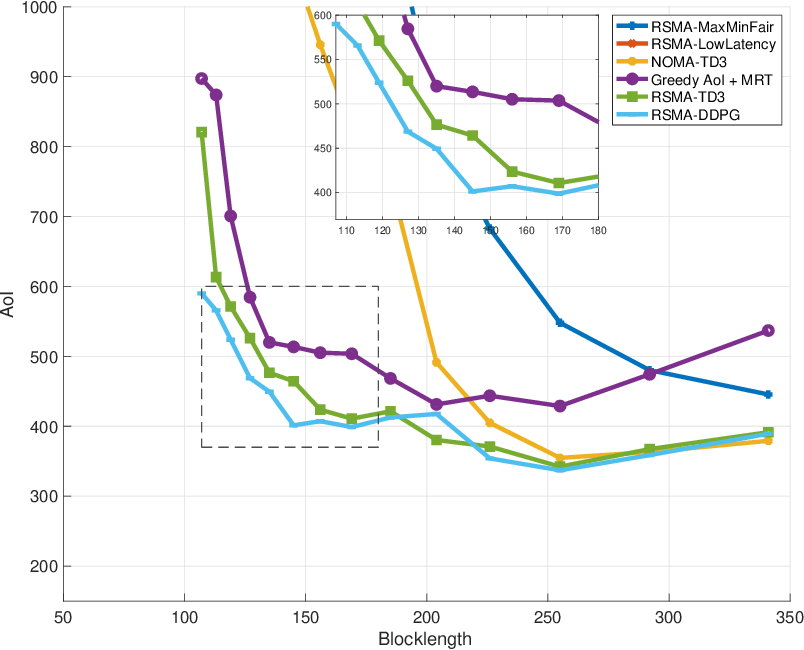}
        \caption{}
        \label{fig:c}
    \end{subfigure}
    \begin{subfigure}{0.40\linewidth}
        \centering
        \includegraphics[width=\linewidth]{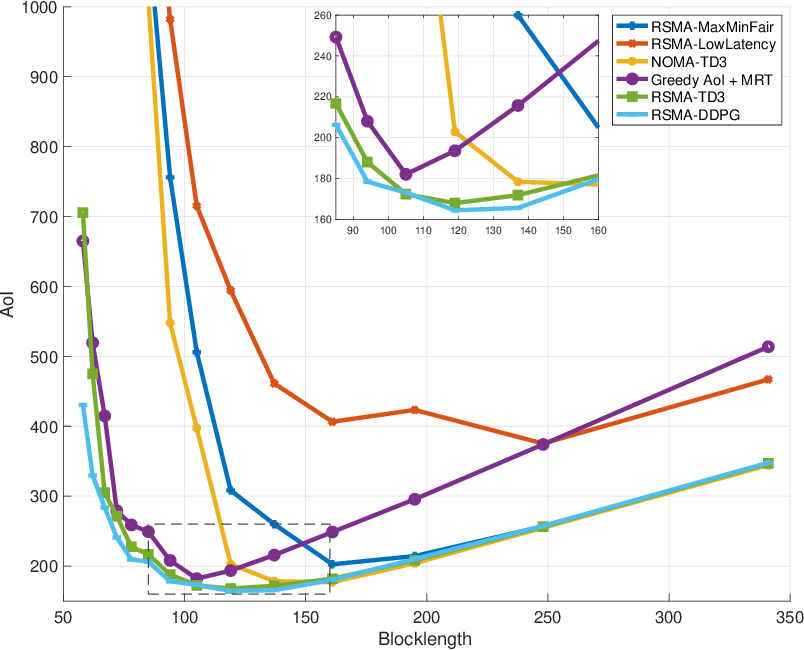}
        \caption{}
        \label{fig:d}
    \end{subfigure}
    \vspace{-0.1in}
    \caption{AoI versus blocklength for a two-antenna base station and single-antenna users, evaluated for (a) 6 dB, perfect CSIT (b) 12 dB, perfect CSIT (c) 6 dB, imperfect CSIT, and (d) 12 dB, imperfect CSIT. Results are averaged over 30 test episodes of 1000 time slots each, following 15000 training episodes of equal length.} 
    \label{fig:mimoperfect}
\end{figure*}

Fig. \ref{fig:mimoperfect}, cases (a) and (b), respectively, illustrate the AoI versus blocklength at SNRs of 6 and 12 dB under perfect channel state information at the transmitter (CSIT), while cases (c) and (d) present the corresponding results under imperfect CSIT. The proposed RSMA-RL frameworks (both TD3 and DDPG) consistently achieve substantially lower AoI than the RSMA-MaxMinFair and RSMA-LowLatency benchmarks, as well as the NOMA-RL method, with particularly pronounced gains in the short blocklength regime. For large blocklengths, the RSMA-RL frameworks significantly outperform the greedy-AoI with MRT benchmark.

At high SNR and long blocklengths, the channel capacity is sufficient to reliably serve both users, reducing resource contention. Consequently, RSMA-MaxMinFair and NOMA-RL become increasingly effective, causing their performance to approach that of the proposed RL-based methods. Nevertheless, the proposed RSMA-RL frameworks can always achieve similar AoI at much shorter blocklengths. 

Among the two RL algorithms, RSMA-DDPG slightly outperforms RSMA-TD3, possibly because TD3's target-policy smoothing is less effective near the sharp FBL decoding threshold. Across all considered scenarios, the proposed RL-based methods exhibit smooth and stable AoI characteristics, with a well-defined U-shaped minimum or a broad, nearly flat minimum that maintains near-optimal AoI over a wide range of blocklengths. This robustness facilitates blocklength selection and reduces sensitivity. In contrast, the benchmarks are considerably more sensitive to the chosen blocklength, with AoI degrading rapidly away from the optimum.

Comparing cases (a) with (b), and (c) with (d), increasing SNR reduces achievable AoI for all schemes and decreases the optimal blocklength; i.e. the minimum AoI decreases from approximately 100 at 6 dB to 60--70 at 12 dB under perfect CSIT, and from about 250 to 110 under imperfect CSIT. Similarly, comparing cases (a) with (c), and (b) with (d), imperfect CSIT increases AoI consistently for all schemes and SNR values. Overall, for the considered two-user setting, the proposed RSMA-RL frameworks consistently outperform all benchmark schemes across the entire SNR range, with the largest gains observed under short-blocklengths, where intelligent and adaptive decision-making is most beneficial.

\vspace{-5pt}
\section{Conclusion}
This letter establishes a novel DRL-based framework for AoI minimization in RSMA-enabled MU-MISO systems under finite-blocklength constraints. The results show that both DDPG and TD3 learn effective dynamic resource-allocation policies that significantly improve timeliness, highlighting DRL's potential for complex optimization in RSMA systems. Future work will investigate scalable extensions based on modular, hierarchical, and multi-agent RL architectures for systems with larger numbers of users and antennas.

\vspace{-5pt}
\bibliographystyle{IEEEtran}
\bibliography{references}

\end{document}